\documentclass[12pt]{article}
\usepackage{amssymb,amsmath,amsthm,url}

\newcommand{\arxiv}[1]{\vphantom{#1}}
\newcommand{\lbr}[2]{ [ \hspace*{-1.6pt} [ #1 , #2 ] \hspace*{-1.7pt} ] }
\newcommand{\inn}{\hspace*{2pt}\raisebox{-1pt}{\rule{6pt}{.5pt}\hspace*
{0pt}\rule{.5pt}{8pt}\hspace*{2pt}}}
\newcommand{\Hgr}{\mathop{\mathrm{Hgr}}}
\newcommand{\comm}[2]{\mathbf{[}#1,#2\mathbf{]}}

\newtheorem{theorem}{Theorem}
\newtheorem{prop}{Proposition}
\theoremstyle{definition}

\title{Poisson bracket in classical field theory \\ as a derived bracket}
\author{S. A. Pol'shin\thanks{E-mail: polshin.s at gmail.com}\\
{\small Institute for Theoretical Physics} \\
{\small NSC Kharkov Institute of Physics and Technology} \\
{\small Akademicheskaia St. 1, 61108 Kharkov, Ukraine }}
\date{}

\begin{document}
\maketitle
\begin{abstract}
We construct a Leibniz bracket on the space $\Omega^\bullet (J^k (\pi))$ of all differential forms over the finite-dimensional jet bundle $J^k (\pi)$. As an example, we write  Maxwell equations with sources in  the covariant finite-dimensional hamiltonian form.



\end{abstract}

The multisymplectic approach to classical field theory  allows us to write the field equations in manifestly covariant finite-dimensional hamiltonian form (see~\cite{GMS97book} for a review) in contrast to the canonical approach based on the 3+1 splitting which is noncovariant and infinite-dimensional. However, there is still a difficulty  with writing the Poisson brackets in the multisymplectic approach (see~\cite{SardBr}). Kanatchikov~\cite{Kan1,Kan2,KanLett} was the first who realized that the notions of \textit{vertical differential} and \textit{Leibniz bracket} are necessary to solve this problem\footnote{Note also the papers Cantrijn, Ibort and de Leon~\cite{Cant} and Forger, Paufler and R\"omer~\cite{Forger}. These authors do not use the vertical differential and handle the narrow class of differential forms only (the so-called hamiltonian forms). On the other hand, the bracket proposed by Grabowski~\cite{Grab} is defined on the whole space of differential forms but fails to satisfy any version of Leibniz rule.}. However, the attentive reader should observe that Kanatchikov's ideas does not fit naturally into the usual multisymplectic framework. Instead, in the present paper we use the Poisson multivector which has the same degree 3 for any number of "space" and "field" dimensions [see eq.~(\ref{multiPois}) below]. Also we systematically use the notion of derived bracket introduced by Y. Kosmann-Schwarzbach~\cite{KS1,KS2} (see also~\cite{CabVin}) who also mentioned that the usual Poisson bracket of functions may be considered on this way. As an example we write Maxwell equations with sources in the Hamiltonian form.

Suppose $\mathcal{D}_V$ and $\mathcal{D}_H$ are two transversal involutive distributions on a smooth manifold $E$, so $TE=\mathcal{D}_H \oplus \mathcal{D}_V$, and let $\mathcal{I}_V$ (resp. $\mathcal{I}_H$) be an ideal in $\Omega^\bullet (E)$ whose elements annihilate distribution $\mathcal{D}_V$ (resp. $\mathcal{D}_H$). Let
$$\Omega^{p,q} =\mathcal{I}_H^p \mathcal{I}_V^q \cap \Omega^{p+q}
(E),$$
where e.g. $\mathcal{I}^p_H$ means the $p$-th power of the ideal $\mathcal{I}_H$. Since $d \mathcal{I}_V  \subset \mathcal{I}_V$ and $d\mathcal{I}_H  \subset \mathcal{I}_H$ due to the Frobenius theorem,  we see that
$$d \Omega^{p,q}\subset \Omega^{p+1,q}\oplus \Omega^{p,q+1},$$
so we can define horizontal and vertical differentials
$d_H :\ \Omega^{p,q}\rightarrow \Omega^{p+1,q}$ and
$d_V :\ \Omega^{p,q}\rightarrow \Omega^{p,q+1}$ such that
$$d^2_H=d_H d_V +d_V d_H=d^2_V=0.$$
We will consider two particular cases.
\begin{enumerate}
\item $E=J^\infty (\pi)$ for some vector bundle $\pi:\ M\rightarrow B$,  $\mathcal{D}_V$ and $\mathcal{D}_H$ are vertical and Cartan distributions respectively (see e.g. Sec.3.1 of~\cite{GMS97book}). We denote the corresponding differentials as $d^{\mathcal{C}}_V$ and $d^{\mathcal{C}}_H$.

\item $E=J^k (\pi)$ for some $k>0$, $\mathcal{D}_V$ is the vertical distribution, $\mathcal{D}_H$ is an orthogonal complementary distribution wrt some pseudo-Riemannian metrics $g$ on $E$ (we assume that the restriction of $g$ onto $\mathcal{D}_V$ is positive-definite). We denote the corresponding differentials as $d^g_V$ and $d^g_H$.
\end{enumerate}

Let $K=\sum\nolimits_{i}^{} K^i$ be a graded module over an arbitrary field $\mathbf{k}$ of characteristics zero and let $\Hgr K=\sum\nolimits_{j}^{}\Hgr^j K$  be a set of all graded
endomorphisms of $K$, $\Hgr^j K:\ K^i\rightarrow K^{i+j}$. Let $F\in \Hgr^{|F|} K$ etc. and let $[\cdot,\cdot]$: $\Hgr^i K \times \Hgr^j K\rightarrow \Hgr^{i+j+f}K$ be a $\mathbf{k}$-bilinear operation on $\Hgr K\times \Hgr K$ satisfying the identities
\begin{gather}
[F,[G,H]]=[[F,G],H]+(-1)^{(|F|+f)(|G|+f)} [G,[F,H]] \label{ident1}\\
[F,G\circ H]=[F,G]\circ H+(-1)^{(|F|+f)|H|} G\circ [F,H] \label{ident2}
\end{gather}
for some fixed $f\in\mathbb{Z}$.
The property~(\ref{ident1}) is an analogue of Jacobi identity (sometimes called the Leibniz one) and~(\ref{ident2}) means that $[\cdot,\cdot]$ is a graded derivation of degree $f$ on the second argument wrt the composition of
endomorphisms of $K$. Note that $[\cdot,\cdot]$ is not necessarily
skew-symmetric. If~(\ref{ident1}) and~(\ref{ident2}) hold then we say that $[\cdot,\cdot]$ is a Leibniz bracket of degree $f$ on $\Hgr K$.

\begin{theorem}[\cite{KS1}]
Let $[\cdot,\cdot]$ be a Leibniz bracket of degree $f$ on $\Hgr K$ and
let $\delta\in \Hgr^{|\delta|} K$ such that $[\delta,\delta]=0$. Then the bracket
$$[F,G]_\delta=[[F,\delta],G]$$
is again a Leibniz bracket of degree $f+|\delta|$ on $\Hgr K$  called the derived bracket of $[\cdot,\cdot]$.
\end{theorem}
For example, the graded commutator $ \comm{F}{G}=F\circ G -(-1)^{|F||G|} G\circ F$ is a Leibniz bracket of degree 0 on $\Hgr K$. In the sequel $[\cdot,\cdot]$ means the graded commutator on $\Hgr K$.

Now let $K=\Omega^\bullet (E)$ and let $\Omega_\bullet (E)$ be a
$\mathbf{k}$-module of smooth  multivector fields on $E$ (this notation is nonstandard). Then it is well known that any $X=X_1 \wedge\ldots\wedge X_n\in \Omega_n (E)$ defines a map $i_X\in \Hgr^n (\Omega^\bullet (E))$ as
$$\Omega^\bullet (E)\ni \alpha\mapsto i_X (\alpha)= X_1 \inn
(\ldots(X_n \inn \alpha))$$
and the map $X\mapsto i_X$ is an injective anti-homomorphism of $\mathbf{k}$-algebras $(\Omega_\bullet (E),\wedge) \hookrightarrow  (\Hgr(\Omega^\bullet (E)),\circ)$ i.e  $i_{X\wedge Y}=i_Y \circ i_X$ and if $i_X (\alpha)=0$ for all $\alpha \in \Omega^\bullet (E)$ then $X=0$.

\begin{prop}
(a) The image of $\Omega_\bullet (E)$ in $\Hgr \Omega^\bullet (E)$ is stable under the derived bracket $\comm{\cdot}{\cdot}_{d^g_V}$, so we can define the vertical Schouten-Nijenhuis bracket $\lbr{\cdot}{\cdot}$ of multivector fields as
$$i_{\lbr{X}{Y}}=\comm{\comm{i_X}{d^g_V}}{i_Y}.$$

(b) If $f\in C^\infty (E)$ and $X\in \Omega_\bullet (E)$, then $\lbr{X}{f}=-\bar{i} (d^g_V f) X$, where $\bar{i}$ is the conjugate insertion operator (see~\cite{Michor,Marle}).

(c) Suppose $X=X_1\wedge\ldots\wedge X_p$ and $Y=Y_1\wedge\ldots\wedge Y_q$ are multivector fields, then
$$\lbr{X}{Y}=\sum\limits_{i,j} (-1)^{i+j} \lbr{X_i}{Y_j}\wedge X_1 \wedge \ldots \hat{X}_i \ldots\wedge X_p\wedge Y_1 \wedge \ldots \hat{Y}_j \ldots\wedge Y_q.$$
\end{prop}
\begin{proof}
Let $X,Y\in \Omega_1 (E)$, then (a) may be easily proved considering a certain trivialization of $\pi$ [see eq.~(\ref{leib-x-y}) below]. On the other hand, using~(\ref{ident1}) we see that $\lbr{X}{Y}=(-1)^{|X|+|Y|+|X||Y|}\lbr{Y}{X}$ since $[i_X,i_Y]=0$. Then for arbitrary $X,Y$ the statement (a) may be proved by induction on their orders using~(\ref{ident2}). The statement (c) may be proved analogously (cf.~\cite{Michor} for the case of ordinary Schouten-Nijenhuis bracket).
\end{proof}

Note that the pseudo-Riemannian metrics $g$ defines an isomorphism $\flat:\ \Omega_1 (E) \rightarrow \Omega^1 (E)$ as $X^\flat (Y)=g(X,Y)$ which can be extended to an isomorphism $(\Omega_\bullet (E),\wedge)\rightarrow (\Omega^\bullet (E),\wedge)$ of exterior algebras due to the functoriality of the exterior algebra construction. The inverse isomorphism we denote as $\sharp$.

Let $P\in \Omega_{|P|} (E)$ be such that $\lbr{P}{P}=0$.
Define the Poisson-Leibniz bracket on $\Omega^\bullet (E)$ as an image of the derived bracket of $\lbr{\cdot}{\cdot}$:
$$\{\alpha,\beta\}_P=\lbr{\lbr{\alpha^\sharp}{P}}{\beta^\sharp}^\flat.$$
Let $s:\ B\rightarrow M$ be a section of $\pi$ and let $j^\infty (s):\ B\rightarrow J^\infty (\pi)$ be the infinite jet of $s$. Consider the equation
\begin{equation}\label{ham}
[j^\infty (s)]^* \left( d^{\mathcal{C}}_H \varphi -\{\chi,\varphi\}_P \right) =0.
\end{equation}
We claim that~(\ref{ham})  defines  hamiltonian evolution corresponding to the potential energy $\chi$, classical observable $\varphi$ and Poisson multivector $P$. Note that we do not need any kinetic terms in $\chi$.

Suppose $\varphi,\psi\in \Omega^\bullet (E)$ obey~(\ref{ham}) for the \textit{same} section $s$ and suppose $P$ is of odd degree. Then it is easily seen that $\varphi\wedge\psi$ obeys~(\ref{ham}) too.

Consider the trivial vector bundle $\pi:\ \mathbb{R}^n \times \mathbb{R}^m \rightarrow \mathbb{R}^n$. Let $x^i$ ($i,j,\ldots=1,\ldots,n$) be coordinates  on $B=\mathbb{R}^n$ and let $u^\alpha$ ($\alpha,\beta,\ldots=1,\ldots,m$) be coordinates on fibers of $\pi$. Then coordinates  of $J^\infty (\pi)$ have the form
$(x^i, u^\alpha_I)$, where $I$ runs over all  sequences of indices $I=(i_1,\ldots,i_l)$  (the empty "sequence" is included as $u^\alpha_{\{\emptyset\}}=u^\alpha$) and $u^\alpha_{\sigma(I)}=u^\alpha_I$ for all permutations of $I$.
On the infinite jets of sections $u^\alpha=s^\alpha (x)$ of $\pi$ we have
\begin{equation}\label{sim}
[j^\infty (s)]^*\, u^\alpha_I    =\frac{\partial^l s^\alpha}
{\partial x^{i_1}\ldots \partial x^{i_l}}.
\end{equation}
Define a certain pseudo-Euclidean metrics on the fibers of $J^k (\pi)\rightarrow B$ for some $k\geq 3$, then $\partial/\partial x^i$, $i=1,\ldots,n$ span the horizontal distibution $\mathcal{D}_H$ and differentials take the form
\begin{gather*}
d^g_V f=\frac{\partial f}{\partial u^\alpha_I}du^\alpha_I, \\
d^{\mathcal{C}}_H f=\left( \frac{\partial f}{\partial x^i}
+\frac{\partial f}{\partial u^\alpha_I}u^\alpha_{I+i} \right)dx^i,
\end{gather*}
where $f\in C^\infty (J^\infty (\pi))$ and $I+i$ is the multiindex obtained by insertion of $i$ into $I$.

It is easily seen that the vertical Schouten-Nijenhuis bracket of vector fields
$$X=X^i\frac{\partial}{\partial x^i}+X^\alpha_I \frac{\partial}{\partial
u^\alpha_I},\qquad Y=Y^j\frac{\partial}{\partial x^j}+Y^\beta_J
\frac{\partial}{\partial u^\beta_J}$$
is equal to
\begin{equation}\label{leib-x-y}
\lbr{X}{Y}=\left(
X^\alpha_I \frac{\partial Y^j}{\partial u^\alpha_I}
 \frac{\partial}{\partial x^j}+
X^\alpha_I \frac{\partial Y^\beta_J}{\partial u^\alpha_I}
 \frac{\partial}{\partial u^\beta_J}\right)-(X\leftrightarrow Y).
 \end{equation}
Consider a multivector $P\in\Omega_3 (E)$ of the form
\begin{equation}\label{multiPois}
   P=\eta^{\alpha\beta} \eta^{IJ}\frac{\partial}{\partial u^\alpha_I}\wedge
\frac{\partial}{\partial u^\beta_{J+i}}\wedge \frac{\partial}{\partial x^i}
\end{equation}
for some constant matrices $\eta^{\alpha\beta}$ and $\eta^{IJ}$. It is easily seen that $\lbr{P}{P}=0$. For the first order theories considered below we put $\eta^{\{\emptyset\}\{\emptyset\}}=1$ and $\eta^{IJ}=0$ unless $I=J=\{\emptyset\}$.

Let $n=1$, then inserting $\varphi=u^\alpha_1$ into the lhs of~(\ref{ham}) we obtain the following equations
which should be valid on sections of $\pi$:
$$u^\alpha_{11}=-\eta^{\alpha\beta}\frac{\partial \chi}{\partial u^\beta}.$$

Put $m=n=4$  (the indices will be denoted as $\mu,\nu,\ldots$) and
$\eta^{\mu\nu}=\mathop{\mathrm{diag}} (+1,+1,+1,-1)$.
 Let $\chi=u^\mu j_\mu /3$ for certain $j^\mu=j^\mu (x)$, then inserting the classical observable
$$\varphi=\frac{1}{2}
\varepsilon^\mu_{\  \nu\rho\sigma} u^\nu_\mu \, dx^\rho \wedge dx^\sigma$$
into lhs of~(\ref{ham}), we obtain
\begin{equation}\label{max}
u^\mu_{\nu\rho}\eta^{\nu\rho}-\eta^{\mu\nu}u^\rho_{\nu\rho}=j^\mu.
\end{equation}
Using~(\ref{sim}) we see that~(\ref{max}) are equivalent to ordinary Maxwell equation with $u^\mu$ being vector-potential. This means that  description of electromagnetic field given above is not gauge-covariant. See also~\cite{Cianci} for the description of gauge fields within the usual multisymplectic approach.


\begin{thebibliography}{99}

\bibitem{GMS97book}
G. Giachetta, L. Mangiarotti  and G. Sardanashvily,
\textit{New Lagrangian and Hamiltonian Methods in Field Theory}
	(World Scientific, Singapore, 1997)

\bibitem{SardBr} G. Sardanashvily, The bracket and the evolution operator in covariant Hamiltonian field theory, math-ph/0209001


\bibitem{Kan1} I. V. Kanatchikov,
Canonical structure of classical field theory in the polymomentum phase space, \textit{Rep. Math. Phys.} \textbf{41}  (1998), 49-90\arxiv{hep-th/9709229}

\bibitem{Kan2} I. V. Kanatchikov,
On field theoretic generalizations of a Poisson algebra,
\textit{Rep. Math. Phys.} \textbf{40}   (1997), 225-234\arxiv{hep-th/9710069}

\bibitem{KanLett} I. V. Kanatchikov,  Novel algebraic structures from the polysymplectic form in field theory, in:  \textit{Physical Applications and Mathematical Aspects of Geometry, Groups and Algebras},  eds. H.-D. Doebner et.al. (World Sci., Singapore, 1997) vol. 2,p. 894\arxiv{hep-th/9712255}

\bibitem{Cant} F. Cantrijn , L. A. Ibort  and M. de Leon,
Hamiltonian structures on multisymplectic manifolds,
\textit{Rend. Semin. Mat. Univ. Pol. Torino} \textbf{54}   (1996), 225-236

\bibitem{Forger}
M. Forger , C. Paufler  and H.  Romer,
The Poisson bracket for Poisson forms in multisymplectic field theory,
\textit{Rev. Math. Phys.} \textbf{15}  (2003), 705-743\arxiv{math-ph/0202043}

\bibitem{Grab} J. Grabowski,
$\mathbb{Z}$-graded extensions of Poisson brackets,
\textit{Rev. Math. Phys.} \textbf{9} (1997), 1-27

\bibitem{KS1} Y. Kosmann-Schwarzbach,
From Poisson algebras to Gerstenhaber algebras,
\textit{Ann. Inst. Fourier (Grenoble)} \textbf{46}   (1996), 1243-1274


\bibitem{KS2} Y. Kosmann-Schwarzbach,
Derived brackets,
\textit{Lett. Math. Phys.} \textbf{69} (2004), 61-87\arxiv{math.DG/0312524}

\bibitem{CabVin} A. Cabras  and A. M. Vinogradov,
Extensions of the Poisson bracket to differential forms and multi-vector fields,  \textit{J. Geom. Phys.} \textbf{9}   (1992), 75-100

\bibitem{Michor}
P. W. Michor,
Remarks on the Schouten-Nijenhuis bracket, \textit{Suppl. Rend. Circ. Mat. Palermo, II. Ser.} \textbf{16}  (1987), 207-215

\bibitem{Marle} Ch.-M. Marle,
The Schouten-Nijenhuis bracket and interior products,  \textit{J. Geom. Phys.} \textbf{23} (1997), 350-359

\bibitem{Cianci} R. Cianci, S. Vignolo and D. Bruno, On the Hamiltonian formulation of Yang--Mills gauge theories, \textit{Int. J. Geom. Meth. Mod. Phys.} \textbf{2} (2005), 1115-1132\arxiv{math-ph/0507001}


\end{thebibliography}
\end{document}